\documentclass[a4paper,11pt]{article}
\usepackage{jcappub} 
\usepackage{lineno}

\usepackage{array} 
\usepackage{booktabs} 
\renewcommand{\arraystretch}{1.3}
\usepackage{hyperref}
\hypersetup{colorlinks=true,linkcolor=blue,citecolor=cyan}
\usepackage{graphicx}
\usepackage{xcolor, soul}
\sethlcolor{green}
\usepackage{dcolumn}
\usepackage{bm}
\usepackage{color}
\usepackage{enumitem}
\usepackage{mathrsfs}
\usepackage{amsmath}
\usepackage{amssymb}
\usepackage{cleveref}
\usepackage{orcidlink}

\crefname{equation}{Eqn.}{Eqns.}
\crefname{figure}{Fig.}{Figs.}
\crefname{section}{Sec.}{Sec.}
\crefname{table}{Table}{Tables}
\usepackage{physics}
\usepackage{multirow}
\setstcolor{red}

\arxivnumber{2503.06750} 
\title{\textcolor{black}{Probing quantum corrected black hole through astrophysical tests\\ with the orbit of S2 star and quasiperiodic oscillations}}

\author[a,b]{Tursunali Xamidov}
\affiliation[a]{Institute of Fundamental and Applied Research, National Research University TIIAME, Kori Niyoziy 39, Tashkent 100000, Uzbekistan} 
\affiliation[b]{Institute for Theoretical Physics \& Cosmology, Zhejiang University of Technology, Hangzhou 310023, China}

\author[a,b,c,d]{Sanjar Shaymatov}
\affiliation[c]{University of Tashkent for Applied Sciences, Str. Gavhar 1, Tashkent 100149, Uzbekistan}
\affiliation[d]{Center for Theoretical Physics, Khazar University, Baku AZ1096, Azerbaijan}

\author[e,f,g]{Bobomurat Ahmedov}
\affiliation[e]{School of Physics, Harbin Institute of Technology, Harbin 150001, People’s Republic of China}
\affiliation[f]{Institute for Advanced Studies, New Uzbekistan University, Movarounnahr str. 1, Tashkent 100000, Uzbekistan}
\affiliation[g]{Institute of Theoretical Physics, National University of Uzbekistan, Tashkent 100174, Uzbekistan}

\author[b,h]{Tao Zhu}
\affiliation[h]{United Center for Gravitational Wave Physics (UCGWP), Zhejiang University of Technology, Hangzhou 310023, China}

\emailAdd{xamidovtursunali@gmail.com}
\emailAdd{sanjar@astrin.uz}
\emailAdd{ahmedov@astrin.uz}
\emailAdd{zhut05@zjut.edu.cn}

 \abstract{
 In this study, we explore the influence of the quantum correction parameter $\xi$ on the motion of particles and the properties of quasiperiodic oscillations (QPOs) around a quantum-corrected black hole (QCBH). We first analyze the geodesics of a test particle and derive weak-field constraints on parameter $\xi$ from the perihelion precession of orbits, using observations from the Solar System and the S2 star's orbit around $\text{SgrA}^\star$ supermassive black hole in the center of our galaxy. We obtain $\xi \leq 0.01869$ and $\xi \leq 0.73528$ using the analysis of Solar System observations and the orbit of the S2 star around $\text{SgrA}^\star$, respectively. In the strong-field regime, we examine the dynamics of epicyclic motion around astrophysical black holes and, using observational data from four QPO sources and the Markov Chain Monte Carlo (MCMC) method, we determine the upper constraint $\xi \leq 2.086$. Our results provide new insights into the effects of quantum corrections on black hole spacetimes and highlight the potential of QPOs as a probe for testing quantum gravity in astrophysical environments.
}

\begin{document}
\maketitle
\flushbottom
\section{Introduction}
\label{introduction}

Albert Einstein's theory of general relativity (GR) revolutionized our understanding of space and time by showing that gravity arises from the curvature of spacetime  \cite{Einstein1916}. This framework provides a robust foundation for describing the macroscopic structure of black holes (BHs), successfully explaining phenomena such as the event horizon, gravitational waves (GWs) \cite{Abbott16a,Abbott16b}, and the imaging of the supermassive black holes (SMBHs) at the center of the $\text{M87}^\star$ \cite{Akiyama19L1, Akiyama19L6} and $\text{SgrA}^\star$ \cite{Akiyama22L12, EHT2022L14} galaxy. The direct detection of gravitational waves, along with the observation and analysis of the supermassive BH image through these instruments, provides explanations for astrophysical phenomena in a strong gravitational field regime. Furthermore, GR explains effects that Newtonian mechanics cannot, including the bending of light by massive objects, the perihelion precession of Mercury, and gravitational redshift. Although GR is a significant and main theory, it does not fully address  singularity problems and quantum nature of spacetime at microscopic scales \cite{Ashtekar:2021kfp}.  This limitation leaves fundamental questions unanswered.  Therefore, testing is required for the development of promising alternative theories that offer a fundamental understanding of these issues. 

The existence of spacetime singularities \cite{Penrose65a, Hawking-Penrose70} pose a fundamental challenge to classical general relativity. A promising way to address this issue involves modifying black hole solutions to include quantum effects. A particularly notable framework for such modifications is the Hamiltonian constraints approach \cite{Ashtekar:2004eh, Thiemann_2007}, which has played a significant role in the canonical quantization of general relativity. Using the Hamiltonian formulation, two black hole models were introduced in \cite{zhang2025blackholescovariance} to address an issue related to general covariance in spherically symmetric gravity, which arises when canonical quantum gravity is applied to develop a semiclassical black hole model. These two models differ in their choice of a quantum parameter. 

It is important to explore whether the quantum parameters can leave any observational signatures for the current/or forthcoming observations, so the quantum effect on the black hole spacetime can be directly tested or constrained by observations. With this motivation, several observational or phenomenological implications of the quantum-corrected black hole spacetime have already been investigated \cite{konoplya2024probingeffectivequantumgravity, shu2024qcbh, skvortsova2024, konoplya2024probingeffectivequantumgravity, Liu:2024soc, Liu:2024wal, Shu:2024tut,Jiang24PDU,Du:2024ujg, Ban:2024qsa, Wang:2024iwt,Uktamov25PDU}. In particular, the light ring and black hole shadow are useful for estimating black hole parameters, including mass, spin, the surrounding external field, and other physical properties. By analyzing the shadows of $\text{M87}^\star$ and $\text{SgrA}^\star$, constraints on the quantum correction parameter $\xi$ have been established \cite{konoplya2024probingeffectivequantumgravity, shu2024qcbh}. In addition, the quasinormal modes of the quantum corrected black hole have also been explored in \cite{skvortsova2024,konoplya2024probingeffectivequantumgravity}.  

The detection of gravitational waves (GWs) and the imaging of the SMBHs at the centers of $\text{M87}^\star$ and $\text{SgrA}^\star$ are pivotal. Therefore, these observations are crucial not only for discovering new solutions in alternative theories of gravity but also for rigorously testing GR in the strong field regime and exploring various gravity theories through precise measurements of spacetime geometry parameters. High-frequency quasi-periodic oscillations (HF QPOs) in X-ray flux from BHs and neutron stars in X-ray binaries offer another promising avenue for testing gravity near BHs. Detected as narrow peaks in power density spectra, these QPOs originate within a few gravitational radii of the central object, depending solely on spacetime geometry and orbital radius. Analyzing QPOs in X-ray data from BH accretion disks \cite{Abramowicz13} can thus reveal valuable insights into spacetime geometry within different gravity theories \cite{Bambi12a,Bambi16b,Tripathi19}. QPOs provide a powerful probe of astrophysical BH candidates in strong gravitational fields, offering valuable insights into the innermost regions of accretion disks and the masses, radii, and spin periods of white dwarfs, neutron stars, and BHs. Consequently, a deeper understanding of the rich nature of QPOs, particularly those described by X-ray power, is essential. This phenomenon has been regularly observed in microquasars, which are considered primary sources of low-mass binary systems (i.e., systems containing neutron stars or BHs). In this context, galactic microquasars are identified as sources of HF QPOs, characterized by upper ($\omega_{U}$) and lower ($\omega_{L}$) frequencies, often exhibiting a 3:2 ratio ($\omega_{U}:\omega_{L}=3:2$) \cite{Kluzniak01,Torok05A&A,Remillard06ApJ}. HF QPO models have been extensively used to study epicyclic motions in BH accretion disks (see Refs.~\cite{Stuchlik13A&A,Stella99-qpo,Rezzolla_qpo_03a}). Notably, HF QPOs typically manifest as twin peaks, with pairs of frequencies observed in most cases. However, the mechanism underlying the occurrence of HF QPOs in specific regions of the accretion disk remains an open question, necessitating the development of explanatory models to address this issue~\cite{Torok11A&A}. Despite this uncertainty, alternative models have been proposed to explain such phenomena, including those incorporating magnetic fields around BHs \cite{Tursunov20ApJ,Panis19,Shaymatov20egb,Shaymatov22c}. In recent years, this topic has spurred significant research activity, with various authors exploring the analysis of HF QPOs within different resonance scenarios and frameworks \cite[see, e.g.,][]{Germana18qpo,Tarnopolski:2021ula,Dokuchaev:2015ghx,Kolos15qpo,Aliev12qpo,Stuchlik07qpo,Stuchlik21:Univ,Titarchuk05qpo,Azreg-Ainou20qpo,Jusufi21qpo,Ghasemi-Nodehi20qpo,Rayimbaev22qpo,Narzilloev23Sym,Shaymatov23ApJ,Shaymatov23qpo,Mustafa24PDU1,Xamidov25PDU, Liu:2023vfh, Guo:2025zca, Liu:2023ggz}.

In this paper, we investigate the effect of the quantum correction parameter $\xi$ on QPOs and the trajectory of particles moving around the QCBH. First, we derive the weak gravitational field constraints on the quantum correction parameter $\xi$ within the observations of the Solar system tests and of the S2 star located in the star cluster close to the $\text{SgrA}^\star$ SMBH using the perihelion precession of particles orbiting the QCBH. Further, enhancing our understanding of the QPOs requires a detailed analysis of the epicyclic motion and the dynamics of timelike particles in the close vicinity of the QCBH.  Next, we apply the Markov Chain Monte Carlo (MCMC) method to determine the best-fit constraints on the parameter $\xi$ in the strong gravitational field regime using observational data from four microquasars - GRO J1655-40, XTE J1550-564, GRS 1915+105, and H1743-322 - collectively referred to as QPO sources.

The paper is organized as follows: In Sec.~\ref{Sec:metric}, we present the spacetime metric of the quantum-corrected black hole (QCBH) and analyze the geodesic motion of a test particle. We derive the weak-field constraints on the quantum correction parameter $\xi$  by studying the perihelion precession of orbits in the Solar System and in the center of the Milky Way galaxy. In Sec.~\ref{Section:EpyFreq}, we investigate the dynamics of epicyclic motion and derive general expressions for the epicyclic frequencies. We further examine how the quantum correction parameter affects these frequencies. Finally, we use the Markov Chain Monte Carlo (MCMC) method in Sec.~\ref{Sec:MCMC} to determine strong-field constraints on parameter $\xi$ using observational data from four QPO sources. We end with a conclusion in Sec.~\ref{Sec:conclusion}.

\section{The spacetime metric and the dynamics of motion}\label{Sec:metric}

In the Hamiltonian formulation of GR, the dynamics are governed by two sets of constraints: the diffeomorphism constraint and the Hamiltonian constraint. The diffeomorphism constraint generates spatial coordinate transformations on each hypersurface, thereby preserving spatial diffeomorphism invariance, while the Hamiltonian constraint governs the evolution of spatial geometry in “time” and ensures the preservation of general covariance.
Recently, the problem of maintaining general covariance in spherically symmetric gravity was examined in Ref.~\cite{zhang2025blackholescovariance}, particularly in the construction of semiclassical black hole models from canonical quantum gravity.  The analysis preserves the kinematical structure of spherically symmetric GR. The study assumes that the classical diffeomorphism constraint remains unchanged, while introducing an effective Hamiltonian constraint, $H_{eff}$, along with an arbitrary function that defines the effective metric. The effective Hamiltonian is defined as follows~\cite{zhang2025blackholescovariance}
\begin{equation}
\begin{aligned}
H_{\text{eff}}&= - \frac{E^{2}}{2 \sqrt{E^{1}}}
- \frac{K_{1} E^{1}}{2 \xi} \sin\!\left( \frac{2 \xi K_{2}}{\sqrt{E^{1}}} \right) - \frac{3 \sqrt{E^{1}} E^{2}}{2 \xi^{2}} 
\sin^{2}\!\left( \frac{\xi K_{2}}{\sqrt{E^{1}}} \right)
+ \frac{K_{2} E^{2}}{2 \xi} \sin\!\left( \frac{2 \xi K_{2}}{\sqrt{E^{1}}} \right) \\
&+ \frac{(\partial_{x} E^{1})^{2}}{8 \sqrt{E^{1}} E^{2}} 
e^{\tfrac{2 i \xi K_{2}}{\sqrt{E^{1}}}}
+ \frac{\sqrt{E^{1}}}{2} \, \partial_{x}\!\left( \frac{\partial_{x} E^{1}}{E^{2}} \right) 
e^{\tfrac{2 i \xi K_{2}}{\sqrt{E^{1}}}} + \frac{i \xi E^{2}}{4} 
\left( \frac{\partial_{x} E^{1}}{E^{2}} \right)^{2} 
\left( \frac{K_{1}}{E^{2}} - \frac{K_{2}}{E^{1}} \right) 
e^{\tfrac{2 i \xi K_{2}}{\sqrt{E^{1}}}} \, ,
\end{aligned}
\end{equation}
where $E^{1}$ and $E^{2}$ are the densitized triad components encoding the spatial geometry, while $K_{1}$ and $K_{2}$ are their conjugate extrinsic-curvature variables. The parameter $\xi$ is the quantum correction parameter.

The effective Hamiltonian constraint deviates from the classical form due to quantum gravity effects.The constraint algebra is assumed to remain the same as in the classical case, but with an additional correction factor $\mu$ that represents quantum gravity effects. As in the classical case, the existence of a Dirac observable associated with the black hole mass is assumed. Based on these assumptions, the authors set up conditions for this observable and derive the corresponding equations that maintain spacetime covariance. The conditions connect the effective Hamiltonian, the Dirac observable for the black hole mass, and the free function. The solution of these equations produces two families of effective Hamiltonian constraints, each characterized by a distinct quantum parameter. In the limit where these parameters vanish, the classical constraints are recovered. Thus, the effective Hamiltonian constraints generate two distinct black hole configurations, each characterized by its own quantum correction parameter.
The spacetime metric describing one of the quantum-corrected black hole (QCBH) is given by \cite{zhang2025blackholescovariance}
\begin{equation} \label{spacetime}
ds^2 = -f(r) dt^2 + f(r)^{-1} dr^2 + r^2 ( d\theta^2 + \sin^2 \theta d\phi^2),
\end{equation}
where 
\begin{equation} 
f(r) = \left(1 - \frac{2M}{r}\right)\left[ 1 + \frac{\xi^2}{r^2} \left( 1 - \frac{2M}{r} \right) \right] ,
\label{metric_funct}
\end{equation}
and $M$ and $\xi$ are the mass of the black hole and the quantum correction parameter, respectively. For further analysis, we shall for simplicity normalize the quantum correction parameter $\xi \to \xi/M$.    
The quantum correction parameter $\xi$ encodes quantum-gravity effects in spacetime, and its normalized form can be written as $\xi = \gamma\sqrt{A}/M$, where $\gamma$ and $A$ are the Barbero–Immirzi  parameter of loop quantum gravity (LQG) and the minimal area gap $A$ associated with holonomy quantization. The area gap, given by $A = 4\sqrt{3}\pi \gamma \ell_{p}^{2}$, corresponds to the smallest non-zero eigenvalue of the area operator in LQG, where $\ell_{p}$ denotes the Planck length~\cite{Ashtekar2006PhRvD}. The parameter $\xi$ indicates the scale at which quantum-gravity effects become more apparent. In the limit $\xi \to 0$, the quantum corrections vanish and the Schwarzschild solution is recovered.
Because the Barbero–Immirzi parameter $\gamma$ is not fixed from fundamental principles in LQG ~\cite{Domagala2004CQG,Meissner2004CQG}, $\xi$ is regarded as a free parameter in our analysis. This allows us to explore possible quantum modifications to black-hole spacetimes arising from LQG.

The Hamiltonian of a neutral particle moving around the QCBH is expressed as follows \cite{Misner73}
\begin{equation}\label{hamiltonian}
H=\frac{1}{2}g^{\alpha\beta}p_\alpha p_\beta \ ,
\end{equation}
where $p^\alpha$ is the four-momentum of the neutral particle. In the spherical coordinate system, $\alpha$ and $\beta$ denote the indices of the $(t, r, \theta, \phi)$ coordinates. The relationship between four-momentum and four-velocity is expressed by the equation $p^\alpha = m u^\alpha,$ where $m$ represents the mass of the neutral particle, $u^\alpha = dx^\alpha/d\tau$ is the four-velocity. Furthermore, $\tau$ defines the proper time. Since the spacetime metric \eqref{spacetime} does not explicitly depend on the $t$ and $\phi$ coordinates, the following conserved quantities can be written \cite{Misner73}
\begin{equation} \label{conservation}
    \frac{p_t}{m}=g_{tt}\frac{dt}{d\tau}=-\mathcal{E}\ ,\qquad
\frac{p_\phi}{m}=g_{\phi\phi}\frac{d\phi}{d\tau}=\mathcal{L}\ ,
\end{equation}
where $\mathcal{E}$ and $\mathcal{L}$ are the specific energy and the specific angular momentum of the particle. Knowing that the Hamiltonian for a massive particle is $H = -m^2/2$ and for a photon is $H = 0$, we can write the following equation using Eq.~\eqref{conservation}

\begin{align} \label{ex-hamiltonian}
   \frac{1}{2}\left(g^{rr}p_r^2+ g^{\theta\theta}p_\theta^2\right)+\frac{m^2}{2}\left( g^{tt}\mathcal{E}^2+ g^{\phi\phi}\mathcal{L}^2\right) = -\frac{m^2}{2} .
\end{align}
After defining $\mathcal{H}$ as:
\begin{equation} \label{Hpot}
    \mathcal{H} = \frac{1}{2} \left( g^{tt} \mathcal{E}^2 + g^{\phi\phi} \mathcal{L}^2 + 1 \right),
\end{equation}
and using 
\begin{eqnarray}
    p_r = g_{rr} p^r = g_{rr} m \frac{dr}{d\tau} , \nonumber \\ 
    p_\theta = g_{\theta\theta} p^\theta = g_{\theta\theta} m \frac{d\theta}{d\tau} \nonumber ,
\end{eqnarray}
 we rewrite Eq.~\eqref{ex-hamiltonian} as:

\begin{equation} \label{eqofmotion}
    g_{rr}\left(\frac{dr}{d\tau}\right)^2+ g_{\theta\theta}\left(\frac{d\theta}{d\tau}\right)^2 = -2\mathcal{H} \ .
\end{equation}

\begin{figure*} 
    \centering
    \includegraphics[scale=0.48]{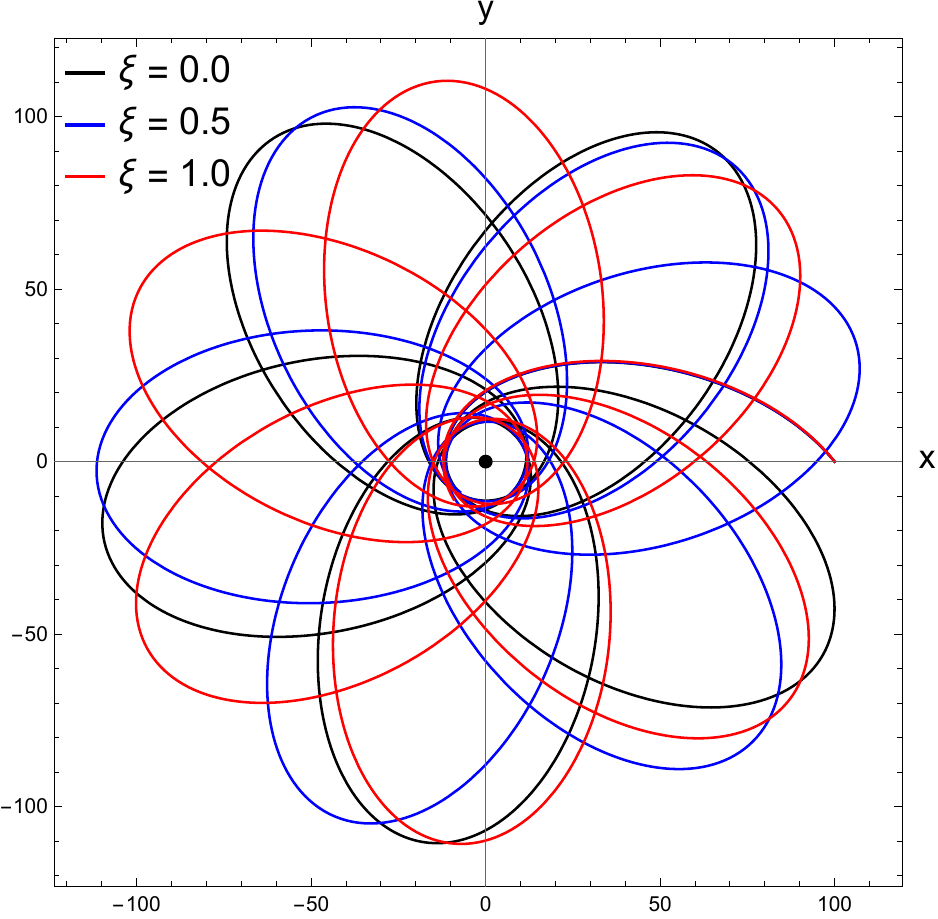}
    \includegraphics[scale=0.48]{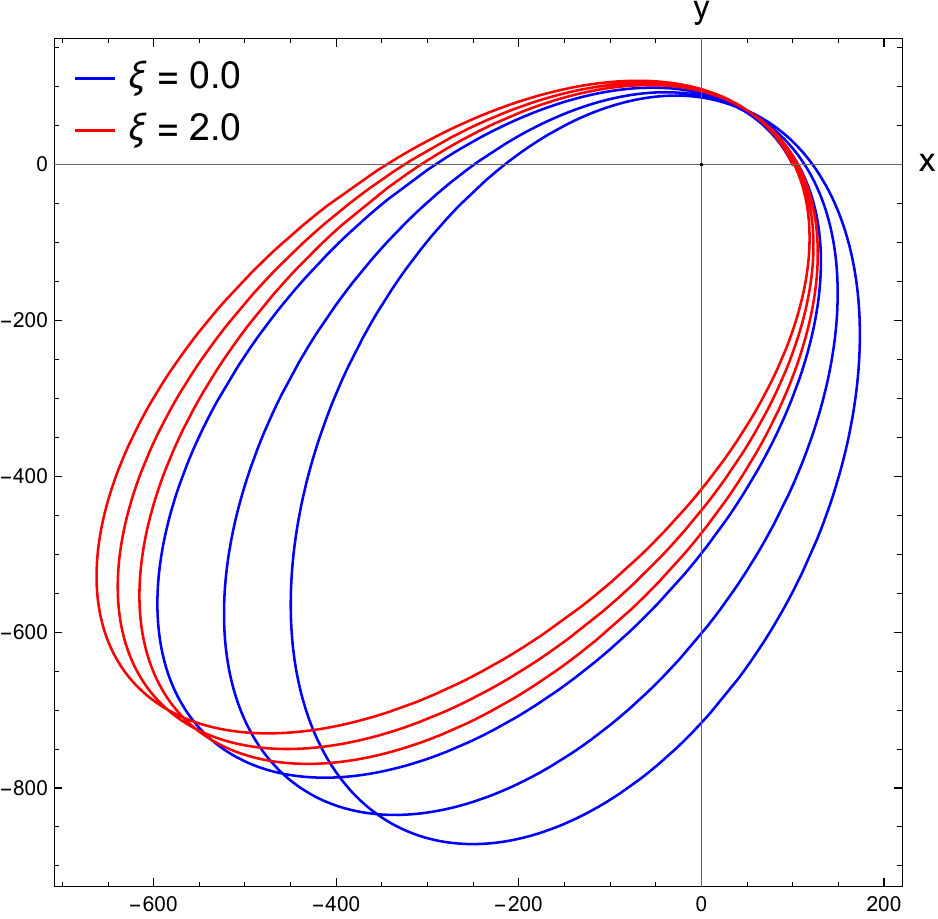}
    
    \caption{The trajectories of a test particle moving in the equatorial plane ($z=0$) of a QCBH spacetime are shown for different values of the quantum correction parameter $\xi$. Left panel: The orbital motion of a particle with specific energy $\mathcal{E} = 0.992$, specific angular momentum $\mathcal{L} = 5$, and initial inverse radial coordinate $u = 1/r = 0.01$ around a black hole of mass $M=1$. Right panel: The effect of $\xi$ on the perihelion shift of a particle with $\mathcal{E} = 0.999$, $\mathcal{L} = 12.5$, and $u = 1/r = 0.01$. As $\xi$ increases, the perihelion shift decreases, highlighting the impact of quantum corrections on the orbital dynamics.
    }

 \label{fig:geodesics}
\end{figure*}
\subsection{The geodesics of a test particle in the equatorial plane}

We consider the motion of a particle confined to the equatorial plane, where $\theta = \pi/2 = const$. Under this assumption, the second term in Eq.~\eqref{eqofmotion} vanishes, and the equation simplifies to the following form:
\begin{equation} \label{equatorial-motion}
    g_{rr}\left(\frac{dr}{d\tau}\right)^2 = -2\mathcal{H} \ .
\end{equation}
Using the conservation of angular momentum Eq.~\eqref{conservation} and Eq.~\eqref{equatorial-motion}, we can derive the following equation of motion:
\begin{eqnarray} \label{equatorial-motion2}
    \left(\frac{dr}{d\phi}\right)^2 = -\frac{2g_{\phi\phi}^2 \mathcal{H}}{g_{rr}\mathcal{L}^2} \, .
\end{eqnarray}
Substituting Eq.~\eqref{Hpot} into Eq.~\eqref{equatorial-motion2} yields the following equation:
\begin{eqnarray} \label{equatorial-motion3}
    \left(\frac{dr}{d\phi}\right)^2 = \frac{r^4}{\mathcal{L}^2}\left(\mathcal{E}^2 - f(r)\left(1+\frac{\mathcal{L}^2}{r^2}\right)\right) \, .
\end{eqnarray}

In the case of the particle moving in a circular orbit $r = const$, the first term of the same equation becomes zero, and the equation takes  the following form
\begin{equation}
    g_{\theta\theta}\left(\frac{d\theta}{d\tau}\right)^2 = -2\mathcal{H} \ .
\end{equation}
The following conditions hold for the circular motion of a particle with the radius $r = r_c$ in the plane $\theta = \theta_c$:

\begin{equation} \label{cond-r}
    \mathcal{H}(r_c) = 0\, , \qquad \partial_r \mathcal{H}|_{r_c} = 0 \, ,
\end{equation}
and
\begin{equation} \label{cond-theta}
    \mathcal{H}(\theta_c) = 0 \, , \qquad \partial_\theta \mathcal{H}|_{\theta_c} = 0 \, .
\end{equation}
If we solve Eq.~\eqref{Hpot} using Eq.~\eqref{conservation} along with conditions~\eqref{cond-r} and~\eqref{cond-theta}, we obtain the following expressions for the specific energy $\mathcal{E}$ and specific angular momentum $\mathcal{L}$

\begin{equation} \label{spec-energy}
\mathcal{E} = -\frac{g_{tt}}{\sqrt{-g_{tt} - g_{\phi\phi}\Omega^2}},
\end{equation}
\begin{equation}\label{spec-ang-momentum}
\mathcal{L} = \frac{g_{\phi\phi}\Omega}{\sqrt{-g_{tt} - g_{\phi\phi}\Omega^2}} \ ,
\end{equation}
where $\Omega = d\phi/dt$ represents the angular velocity of the particle measured by distant observers. It can be calculated using the following expression \cite{Shaymatov22a,Shaymatov23qpo}:
\begin{equation} \label{omega}
\Omega =\frac{d\phi}{dt}=\pm\sqrt{ -\frac{g_{tt,r}}{g_{\phi\phi,r}}} \, .
\end{equation}
We can easily define $\Omega$ from the metric \eqref{spacetime} using the above expression as 

\begin{equation}
    \Omega =\pm\sqrt{\frac{M r^3-\xi ^2 \left(8 M^2-6 M r+r^2\right)}{r^6}\csc ^2(\theta )} \ .
\end{equation}
Thus, the following expressions for $\mathcal{E}$ and $\mathcal{L}$ can be written
\begin{equation} \label{spec-energy-exp}
\mathcal{E} =\frac{(2 M-r) \left(2 M \xi ^2-r \left(\xi ^2+r^2\right)\right)}{r^2 \sqrt{(r-3 M) \left(2 \xi ^2 (r-2 M)+r^3\right)}},
\end{equation}

\begin{equation}\label{spec-ang-momentum-exp}
\mathcal{L} = \sqrt{\frac{r^2 \left(M r^3-\xi ^2 \left(8 M^2-6 M r+r^2\right)\right)}{(r-3 M) \left(2 \xi ^2 (r-2 M)+r^3\right)} \sin ^2(\theta )} \, .
\end{equation}
After making the following transformation
\begin{eqnarray}
    u = \frac{1}{r} , \quad \frac{du}{d\phi} = -\frac{1}{r^2}\frac{dr}{d\phi} \, ,
\end{eqnarray} 
Eq.~\eqref{equatorial-motion3} becomes:
\begin{eqnarray}
\label{eqGeodesic}
    \left(\frac{du}{d\phi}\right)^2 = \frac{\mathcal{E}^2}{\mathcal{L}^2}- \frac{f(u)}{\mathcal{L}^2}\left(1+\mathcal{L}^2u^2\right) \, ,
\end{eqnarray}
where
\begin{equation}
    f(u) = (1-2 M u)(1+\xi^2 u^2(1-2 M u))\, .
\end{equation}
By differentiating Eq.~\eqref{eqGeodesic} with respect to $\phi$, we obtain the geodesic equation for a massive test particle moving around the quantum corrected black hole
\begin{align} \label{orbital-equation}
    \frac{d^2 u}{d\phi^2} = \frac{M}{\mathcal{L}^2} - u+ 3 M u^2 
    -   \xi^2 u ( 1-2 M u) \left( \frac{1-4 M u}{\mathcal{L}^2}+ 2  u^2-6 M u^3   \right).
\end{align}
\subsection{The perihelion shift}
To determine the perihelion shift of the test particle's orbit, we rewrite Eq.~\eqref{orbital-equation}
\begin{align} \label{orbital-equation2}
    &\frac{d^2 u}{d\phi^2} =  \frac{M}{\mathcal{L}^2}- u + \frac{g(u)}{\mathcal{L}^2} \, , 
\end{align}
where
\begin{equation*}
    \frac{g(u)}{\mathcal{L}^2} = 3 M u^2 
    -   \xi^2 u ( 1-2 M u) \left( \frac{1-4 M u}{\mathcal{L}^2}+ 2  u^2-6 M u^3   \right)\, .
\end{equation*}
Following the method presented in \cite{Adkins_2007}, we define  the perihelion shift after a full revolution as:
\begin{equation}
    \Delta \varphi = \frac{\pi}{\mathcal{L}^2} \left| \frac{d g(u)}{du} \right|_{u = \frac{1}{b}}\, ,
\end{equation}
where $b$ is defined as $b = a(1-e^2)$. Here, $a$ and $e$ denote the semi-major axis and eccentricity of the orbit, respectively. Additionally, we define $M$ as $M \Rightarrow GM/c^2$, and $L^2$ as $L^2 \Rightarrow GM a(1-e^2)/c^2$ in SI units. After this transformation, the expression takes the following form:
\begin{eqnarray} \label{eq:perishift}
    \Delta\phi = \frac{6 \pi  G M}{a c^2 (1-e^2)}-\frac{\pi  G M\xi^2}{a c^2 (1-e^2) } \Bigg(1- \frac{6   G M}{a c^2 (1-e^2) }-
    \frac{16   G^2 M^2}{a^2 c^4 (1-e^2)^2 }+\frac{60   G^3 M^3}{a^3 c^6 (1-e^2)^3 }\Bigg)\, .
\end{eqnarray}
For a numerical application, we consider the example of Mercury and use the following parameters
\begin{eqnarray*}
    \frac{2 G M_\odot}{c^2} &=& 2.95325008 \times 10^3 \, [\text{m}] \, , \nonumber \\
    a &=&  5.7909175 \times 10^{10} \, [\text{m}] \, , \nonumber \\
    e &=&  0.20563069\, .
\end{eqnarray*}
With these parameters, we derive the following numerical result for Mercury's perihelion shift
\begin{equation} \label{numShift}
    \Delta\phi = 2\pi\times(7.98744\times 10^{-8})-2\pi\times(1.33124\times10^{-8})\xi^2 \, .
\end{equation}
The first and second terms correspond to the prediction of general relativity and the influence of the quantum correction parameter, respectively. The measured perihelion shift of Mercury is expressed as \cite{Benczik02PRD,Iorio15IJMPD,Iorio2019ApJ,Shaymatov23ApJ}
\begin{equation}\label{obsShift}
    \Delta \phi_{\text{obs}} = 2\pi \times (7.98734 \pm 0.00037) \times 10^{-8} \text{ rad/rev} \, .
\end{equation}
Using Eq.~\eqref{numShift} and Eq.~\eqref{obsShift}, we can estimate the possible value of the quantum correction parameter $\xi$ for Mercury
\begin{equation} \label{xi-mercury}
    \xi \leq 0.01869 \, .
\end{equation}
The influence of metric-affine geometry on particle motion becomes significant in strong gravitational regimes. One such scenario occurs in high-density astrophysical environments, such as near a supermassive black hole. High-precision astrometric observations near the compact source $\text{Sgr A}^*$ and the S2 star, which orbits this supermassive black hole and has an estimated mass between $10$ and $15 M_{\odot}$, allow us to estimate the quantum correction parameter $\xi$ \cite{AbuterAmorim2020}. In general, the rotation of a black hole can influence the dynamics of stellar stars, such as the S2 star, orbiting a supermassive black hole. However, in this study, we neglect the effect of black hole rotation, as its impact on the motion of the S2 star is negligibly small. The S2 star can be treated as a test body moving along a geodesic trajectory. For this estimation, we adopt the following parameters:
\begin{eqnarray*}
    G &=& 6.6743 \times 10^{-11} \, [\text{m}^3/(\text{kg} \cdot \text{s}^2)]\, , \\
    c &=& 299,792,458 \, [\text{m/s}]\, ,\\
    M_{\odot} &=& 1.988416 \times 10^{30} \, [\text{kg}] \, , \\
    M_{\text{Sgr A}^*} &=& 4.260 \times 10^6 M_{\odot}\, , \\
    a_{\text{S2}} &=& 970 \, [\text{au}] \, , \\
    1 \ \text{au} &=& 1.495978707\times10^{11} \, [\text{m}]\, , \\
    e_{\text{S2}} &=& 0.884649 \, , \\
    T_{\text{S2}} &=& 16.052 \, [\text{years}] \, .
\end{eqnarray*}
Using these parameters, we obtain the following numerical result for the perihelion shift of the S2 star
\begin{equation} \label{numShiftS2}
    \Delta\phi = 48.298 -8.040\ \xi^2 \Big[\ ^{\prime\prime}/\text{year} \Big] \, .
\end{equation}
The observed perihelion shift of the S2 star orbiting $\text{Sgr A}^*$ is given by
\begin{equation} \label{obsShiftS2}
    \Delta\phi = 48.298 \ f_{\text{SP}} \ \Big[\ ^{\prime\prime}/\text{year} \Big] \, ,
\end{equation}
where $f_{\text{SP}} = 1.10 \pm 0.19$  \cite{AbuterAmorim2020}. By utilizing Eqs.~\eqref{numShiftS2} and \eqref{obsShiftS2}, we can determine the possible value of the quantum correction parameter $\xi$ for the S2 star
\begin{equation} \label{xi-S2}
    \xi \leq 0.73528 \, .
\end{equation}
Fig.~\ref{fig:geodesics} illustrates the effect of the quantum correction parameter $\xi$ on the trajectory of a particle orbiting a QCBH. The left panel of  Fig.~\ref{fig:geodesics} illustrates the trajectory of a particle for different values of $\xi$, while keeping the initial conditions energy, angular momentum, and initial position constant. The black trajectory represents the classical motion of a particle orbiting a Schwarzschild black hole ($\xi = 0.0$). As $\xi$ increases, the trajectories deviate significantly from the classical case, demonstrating the impact of quantum corrections on the particle’s motion.
The right panel of Fig.~\ref{fig:geodesics} depicts the effect of the quantum correction parameter on the perihelion shift of the particle. The graph shows that as $\xi$ increases, the perihelion shift decreases. This trend is in agreement with the analytical result given by Eq.~\eqref{eq:perishift}, confirming that quantum corrections reduce the relativistic precession of the orbit.

\section{The dynamics of epicyclic motion}
\label{Section:EpyFreq}
Suppose that a particle is moving in a circular orbit of radius $r = r_c$ in the equatorial plane $\theta = \pi/2$. If the particle deviates from its stable circular orbit by small perturbations $\delta r$ and $\delta\theta$, it begins to oscillate around the circular orbit at $r_c$. These oscillations, characterized by radial and latitudinal frequencies, are collectively referred to as epicyclic frequency. We can write the following expressions for $r$ and $\theta$ for this perturbations:
\begin{eqnarray} \label{perturb-r-th}
    r = r_c + \delta r \quad \text{and} \quad \theta = \pi/2 + \delta\theta \, .
\end{eqnarray}
The change in $\mathcal{H}$ due to this perturbation can be determined by expanding $\mathcal{H}$ in a Taylor series about $r = r_c$ and $\theta = \pi/2$

\begin{align} \label{h-taylor}
    &\mathcal{H}(r,\theta) = \mathcal{H}(r_c,\frac{\pi}{2}) + \delta r \partial_r \mathcal{H}(r,\theta)\Big|_{r_c,\frac{\pi}{2}}  + \delta \theta \partial_\theta \mathcal{H}(r,\theta)\Big|_{r_c,\frac{\pi}{2}}+\delta r\delta\theta\partial_r \partial_\theta \mathcal{H}(r,\theta)\Big|_{r_c,\frac{\pi}{2}} +\nonumber\\ & + \frac{1}{2}\delta r^2 \partial_r^2 \mathcal{H}(r,\theta)\Big|_{r_c,\frac{\pi}{2}} + \frac{1}{2}\delta \theta^2 \partial_\theta^2 \mathcal{H}(r,\theta)\Big|_{r_c,\frac{\pi}{2}} + \mathcal{O}\left( \delta r^3, \delta \theta^3 \right) \ .
\end{align}
According to conditions~\eqref{cond-r} and~\eqref{cond-theta} for circular motion, the first four terms in the above expression vanish. Additionally, since they are very small, terms of $\mathcal{O}(\delta r^3, \delta \theta^3)$ and higher can be neglected. As a result, the expression simplifies to the following form:
\begin{align} \label{h-taylor}
    \mathcal{H}(r,\theta) =  \frac{1}{2}\delta r^2 \partial_r^2 \mathcal{H}(r,\theta)\Big|_{r_c,\frac{\pi}{2}} + \frac{1}{2}\delta \theta^2 \partial_\theta^2 \mathcal{H}(r,\theta)\Big|_{r_c,\frac{\pi}{2}} \ .
\end{align}
Substituting expressions~\eqref{perturb-r-th} and~\eqref{h-taylor} into Eq.~\eqref{eqofmotion}, we obtain the following equation:

\begin{align} \label{eq-mo-h-t}
    &g_{rr} \delta\dot{r}^2+ g_{\theta\theta}\delta\dot{\theta}^2  = -\delta r^2 \partial_r^2 \mathcal{H}(r,\theta)\Big|_{r_c,\frac{\pi}{2}} -\delta \theta^2\partial_\theta^2 \mathcal{H}(r,\theta)\Big|_{r_c,\frac{\pi}{2}} \ .
\end{align}
Here, since $d$ and $\delta$ commute, we have made the substitutions $d\delta r/d\tau = \delta\dot{r}$ and $d\delta\theta/d\tau=\delta\dot{\theta}$. Taking the first derivative of both sides of the above equation with respect to proper time $\tau$, we obtain 
\begin{align} \label{sim-eq-mo}
    &\delta\dot{r}\Big[ g_{rr} \delta\ddot{r}+ \partial_r^2 \mathcal{H}(r,\theta)\Big|_{r_c,\frac{\pi}{2}} \delta r\Big]  + \delta\dot{\theta}\Big[ g_{rr} \delta\ddot{\theta}+ \partial_\theta^2 \mathcal{H}(r,\theta)\Big|_{r_c,\frac{\pi}{2}} \delta \theta\Big] = 0 \ .
\end{align}
The solutions $\delta\dot{r} = 0$ and $\delta\dot{\theta} = 0$ are not satisfactory, as they indicate the absence of oscillations. In this case, the equation has a solution only if the expressions within the brackets are zero \cite{Shaymatov20egb,Stuchlik21_qpo}
\begin{equation}
    \delta \ddot{r} + \Omega_r^2 \delta r = 0 \quad \text{and} \quad \delta \ddot{\theta} + \Omega_\theta^2 \delta \theta = 0 \, , 
\end{equation}
where
\begin{equation}\label{radial-freq}
    \Omega_r^2 = \frac{\partial_r^2 \mathcal{H}(r,\theta)|_{r_c,\frac{\pi}{2}}}{g_{rr}} \ ,
\end{equation}
\begin{equation} \label{latitudal-freq}
    \Omega_\theta^2 = \frac{\partial_\theta^2 \mathcal{H}(r,\theta)|_{r_c,\frac{\pi}{2}}}{g_{\theta\theta}} \ .
\end{equation}
By substituting Eq.~\eqref{spec-energy-exp}, \eqref{spec-ang-momentum-exp} and \eqref{Hpot} into Eq.~\eqref{radial-freq} and ~\eqref{latitudal-freq}, we obtain the following expressions for $\Omega_r$ and $\Omega_\theta$
\begin{align} \label{radial-freq-exp}
    &\Omega_r^2 =\frac{3 M \xi ^2 r^3 \left(12 M^2-8 M r+r^2\right)+M r^6 (r-6 M)}{r^6 (r-3 M) \left(2 \xi ^2 (r-2 M)+r^3\right)} - \nonumber \\ 
    & - \frac{2 \xi ^4 (r-2 M)^2 \left(24 M^2-13 M r+2 r^2\right)}{r^6 (r-3 M) \left(2 \xi ^2 (r-2 M)+r^3\right)} \ ,
\end{align}

\begin{equation} \label{latitudal-freq-exp}
    \Omega_\theta^2 =\frac{M r^3-\xi ^2 \left(8 M^2-6 M r+r^2\right)}{r^2 (r-3 M) \left(2 \xi ^2 (r-2 M)+r^3\right)} \ .
\end{equation}

\begin{figure}
    \centering
    \includegraphics[scale=0.55]{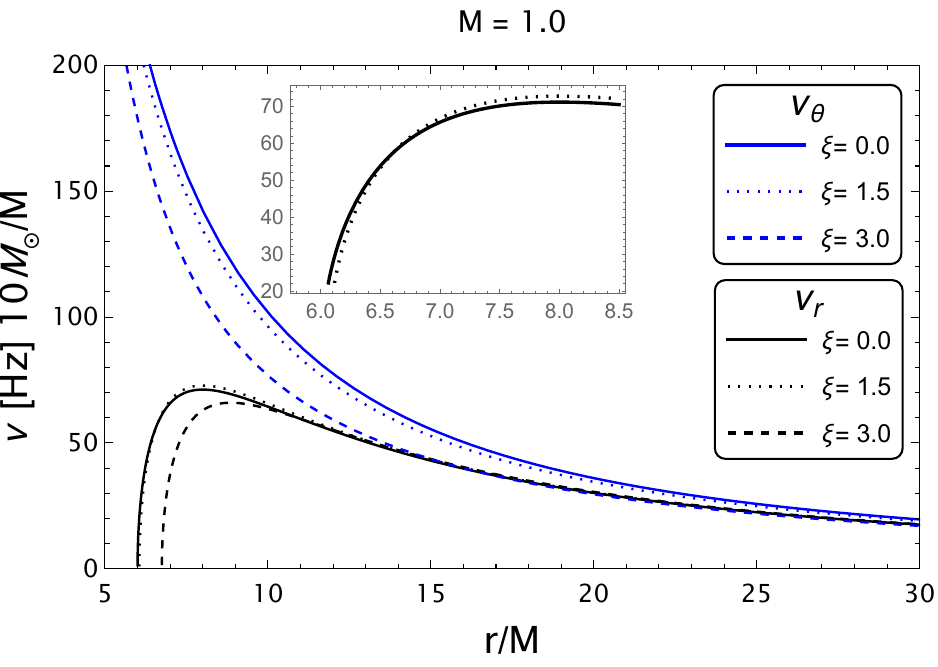}
    \caption{The radial profile of the frequencies $\nu_\theta$ and $\nu_r$ measured by a distant observer is shown as a function of $r/M$ for different values of the quantum correction parameter $\xi$. 
 }
    \label{fig:frequency}
\end{figure}
Usually, it is not the local value of the epicyclic frequency that is important but the value measured by a distant observer, as many astronomical objects are located far away. The following equation gives the relationship between local and distant epicyclic frequencies: 
\begin{equation}
    \omega_{distant} = \frac{\Omega_{local}}{u^t} \ ,
\end{equation}
where $u^t = dt/d\tau$, which can be determined by substituting Eq.~\eqref{spec-energy} into Eq.~\eqref{conservation}
\begin{equation}\label{four-vel}
u^t = \frac{1}{\sqrt{-g_{tt} - g_{\phi\phi}\Omega^2}} \ .
\end{equation}
Thus, the expression for the epicyclic frequencies measured by a distant observer is given as follows:
\begin{align} \label{radial-freq-exp-distant}
    &\omega_r^2 =\frac{3 M \xi ^2 r^3 \left(12 M^2-8 M r+r^2\right)+M r^6 (r-6 M)}{r^{10}} - \nonumber \\ 
    & - \frac{2 \xi ^4 (r-2 M)^2 \left(24 M^2-13 M r+2 r^2\right)}{r^{10}} \ ,
\end{align}
\begin{figure*}
    \centering
    \includegraphics[scale=0.48]{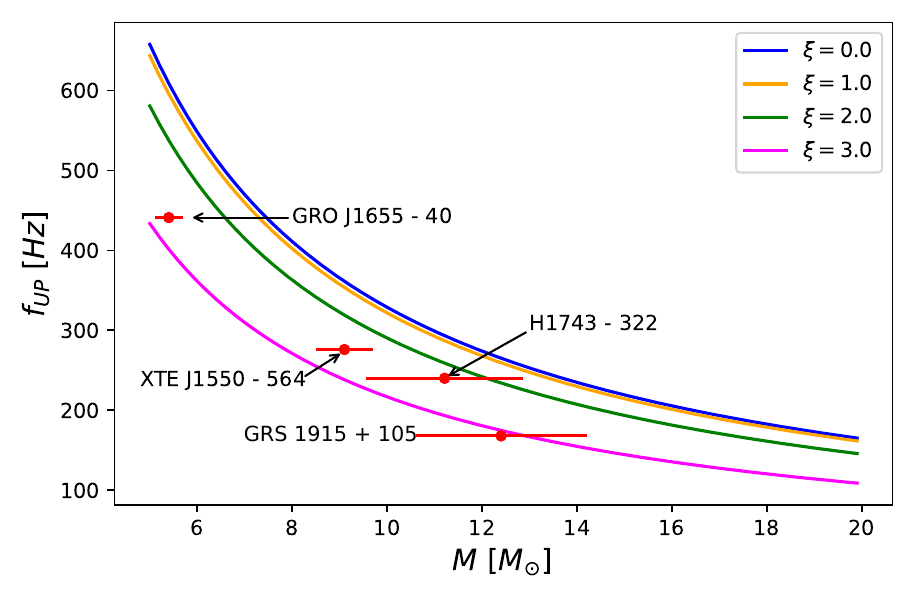}
    \includegraphics[scale=0.48]{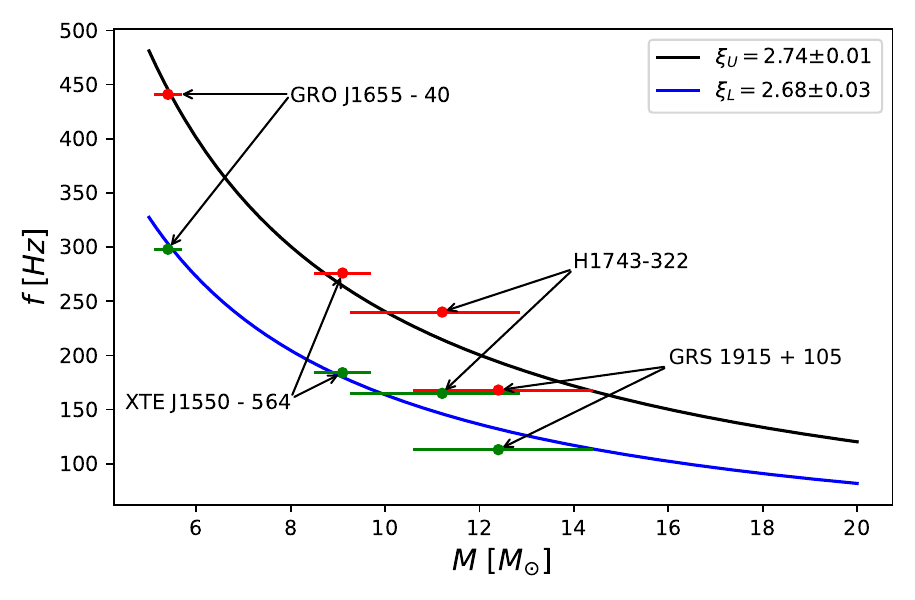}
    
    \caption{Left: The upper $\nu_U$ frequency is depicted as functions of the mass $M$ for different values of the quantum correction parameter $\xi$, analyzed near the innermost stable circular orbit $r_{\text{isco}}$. Red horizontal lines indicate the observational data for various QPO sources, with the width of these lines representing the uncertainty in the mass of the sources. Right: The upper and lower frequencies of a particle near the innermost stable circular orbit $r_{\text{isco}}$ are plotted as function of $M/M_\odot$ for the best-fit values of the quantum correction parameter $ \xi $. The black solid curve represents the best-fit curve of the upper frequency $ f_U $, while the blue solid curve corresponds to the lower frequency $f_L$. Observational data for various QPO sources are shown as horizontal lines, where red lines indicate the upper frequency and green lines represent the lower frequency. The width of these horizontal lines reflects the uncertainty in the mass estimates of the sources.
 }
    \label{fig:freq-UP}
\end{figure*}
\begin{equation} \label{latitudal-freq-exp-distant}
    \omega_\theta^2 =\frac{M r^3-\xi ^2 \left(8 M^2-6 M r+r^2\right)}{r^6 } \ .
\end{equation}
To express the frequency in SI units (Hz), the following transformation is required
\begin{equation}
    \nu_i = \frac{\omega_i}{2\pi} \frac{c^3}{GM} \ .
\end{equation}

\begin{figure*}
\includegraphics[scale=0.4]{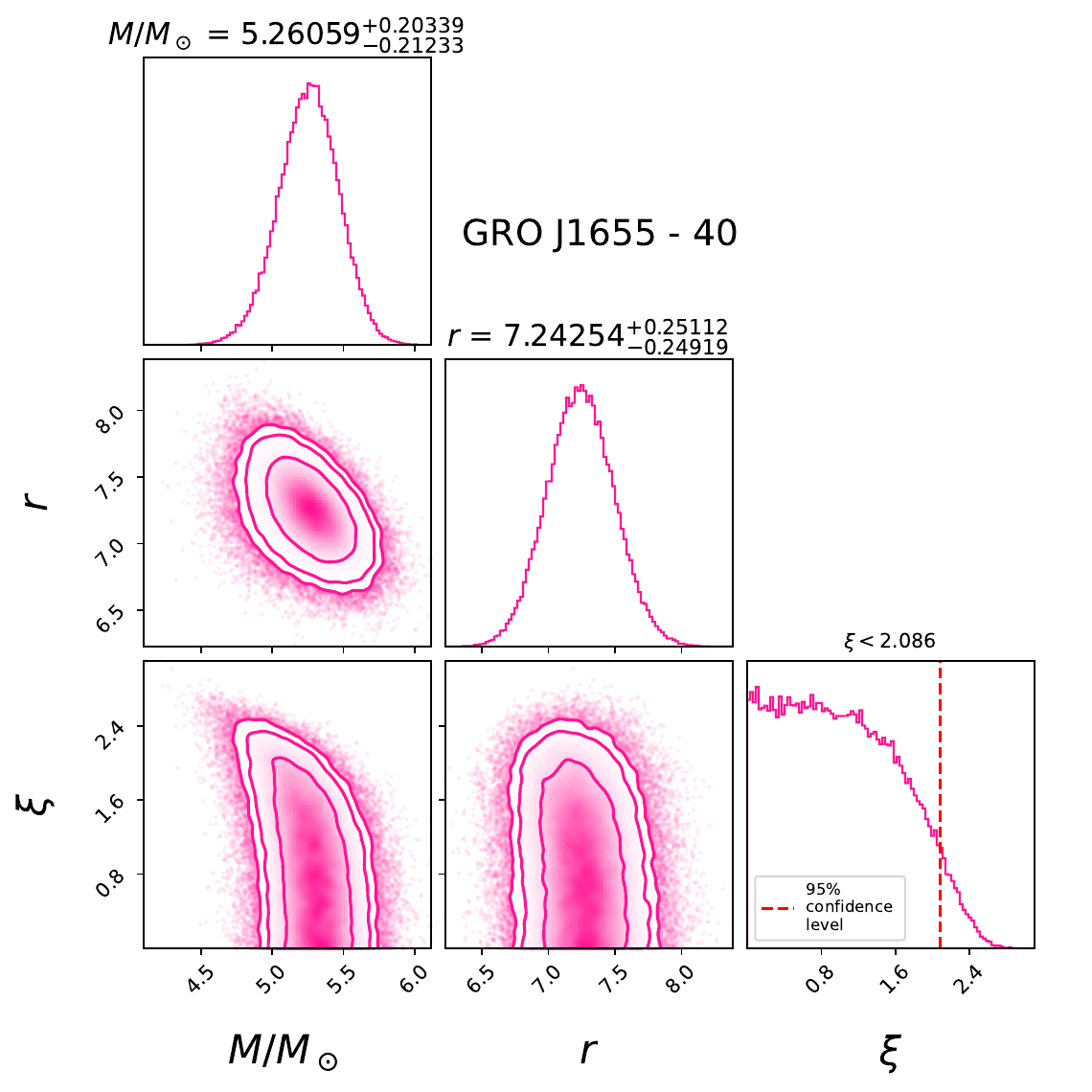}
\includegraphics[scale=0.4]{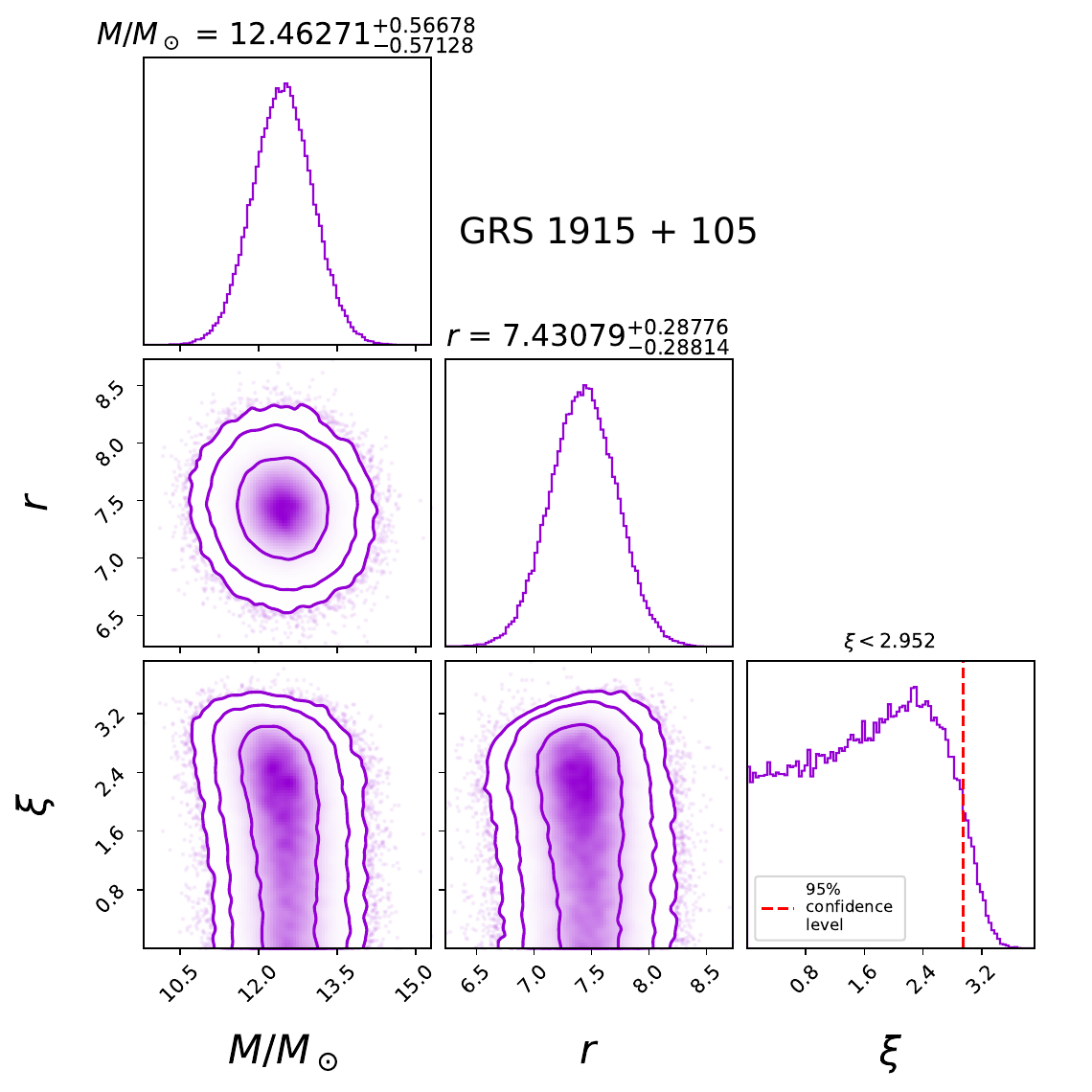}
\includegraphics[scale=0.4]{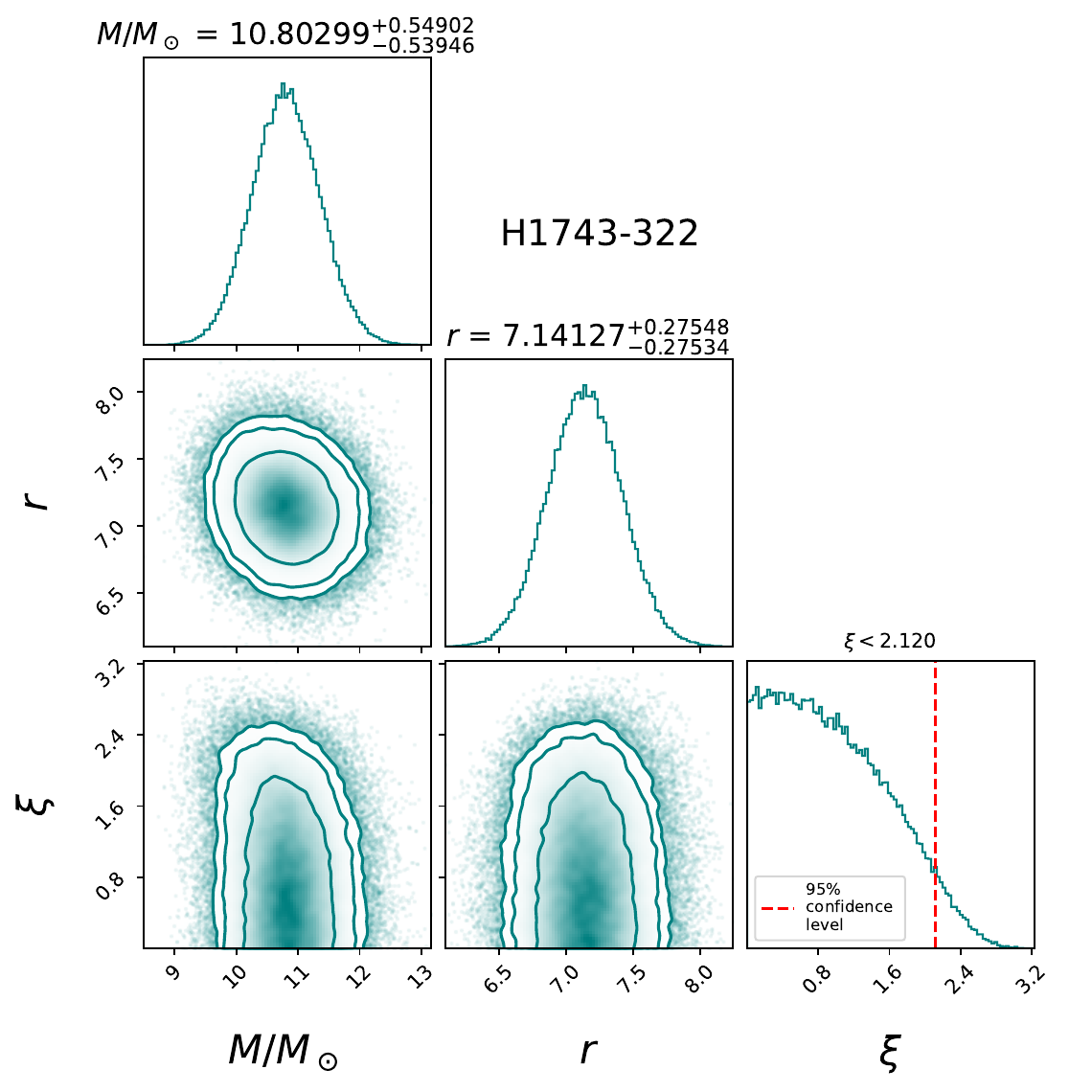}
\includegraphics[scale=0.4]{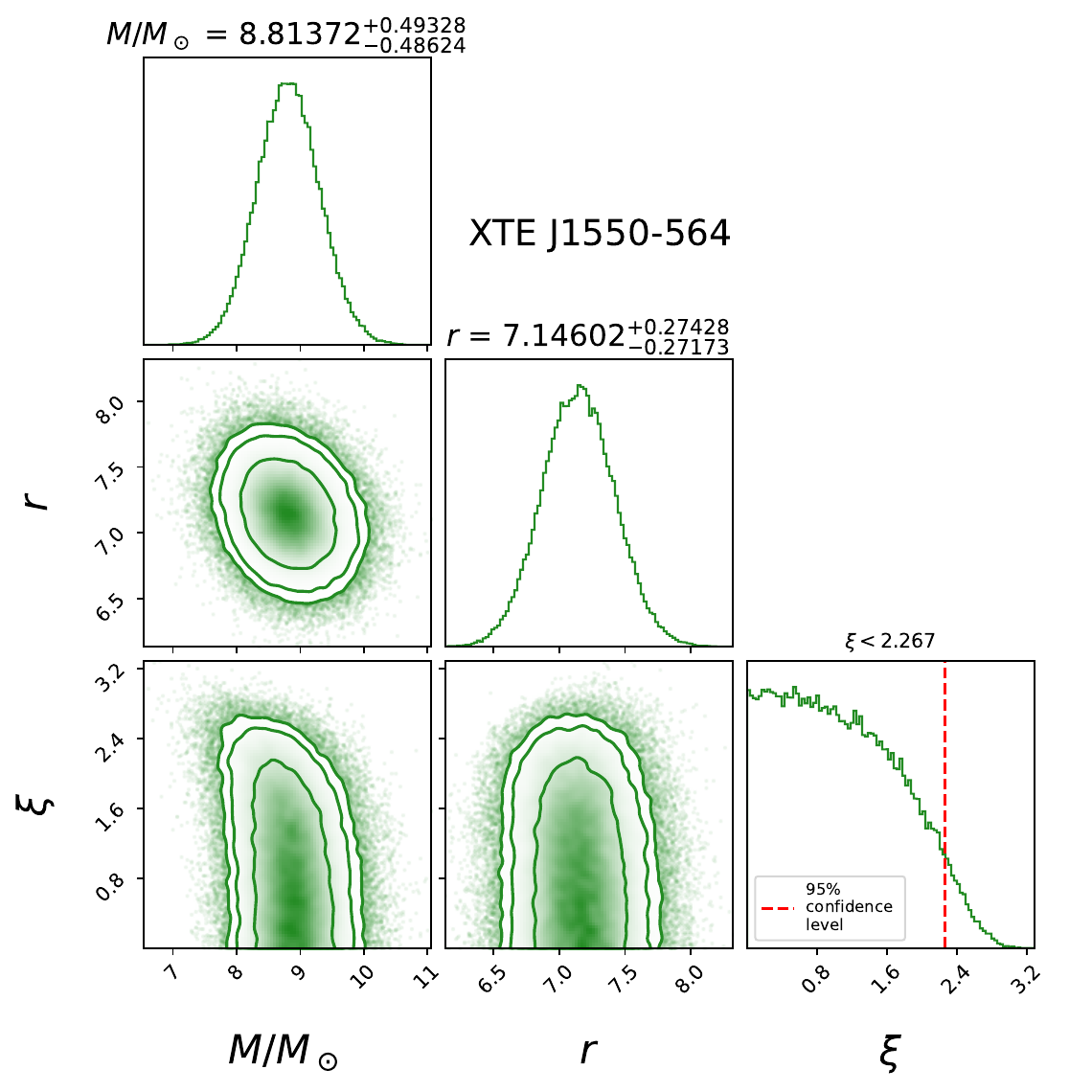}    
    \caption{Constraints on the parameters of QCBH from QPO observations of GRO J1655-40, GRS 1915+105, H1743-322, and XTE J1550-564 using the MCMC method. The plots show the posterior distributions for the black hole mass $ M $, the dimensionless radius $ r/M $, and the quantum correction parameter $ \xi $ within the forced resonance mode. The vertical dashed red lines indicate the 95\% confidence level for $ \xi $.}
    \label{Fig.mcmc}
\end{figure*}
Fig.~\ref{fig:frequency} presents the epicyclic frequencies $\nu_r$ and $\nu_\theta$ as a function of $r/M$, as measured by a distant observer, for different values of the quantum correction parameter $\xi$. Both $\nu_r$ and $\nu_\theta$ decrease as $r/M$ increases. Higher values of the quantum correction parameter $\xi$  result in lower oscillation frequencies, shifting $\nu_\theta$ to the left and $\nu_r$ to the right.

\section{Constrains on the parameters of the quantum corrected black hole through the astrophysical quasiperiodic oscillations's data }\label{Sec:MCMC}
In this section, we analyze four distinct QPO sources from various X-ray binaries, using validated observational data, as presented in Table~\ref{Table 1} to constrain the effects of the quantum correction parameter $\xi$ in QCBH. 
There exist several leading models proposed to explain the origin of QPOs, including the relativistic precession (RP), epicyclic resonance (ER), tidal disruption (TD), and forced resonance models.
Here, we adopt the forced resonance model as the theoretical framework to constrain the quantum correction parameter $\xi$ of QCBH \cite{Deligianni2021,Banerjee_2022,Shaymatov23ApJ}. Within this model, the upper (\(\nu_U\)) and lower (\(\nu_L\)) frequencies are defined as follows.
\begin{equation} \label{upper-lower-freq}
    \nu_U = \nu_\theta + \nu_r\, , \, \nu_L = \nu_\theta \ .
\end{equation}
In Fig.~\ref{fig:freq-UP}, we plot the upper and lower frequencies of a particle near $r_{\text{isco}}$ as a function of the black hole mass $M$ for various values of $\xi$. The left panel of the figure illustrates the effect of $ \xi $ on the upper frequencies. Clearly, the upper frequency $f_{\text{UP}}$ decreases as both $\xi$ and $ M $ increase. As the mass $M$ increases, the gravitational force that pulls the particle toward the black hole also becomes stronger. Consequently, the innermost stable circular orbit expands, causing the QPO to occur farther from the black hole. As a result, the QPO frequency decreases, as indicated by Eqs.~\eqref{radial-freq-exp} and \eqref{latitudal-freq-exp}. In the right panel of Fig.~\ref{fig:freq-UP}, the upper and lower frequencies of a particle near the innermost stable circular orbit $ r_{\text{isco}} $ are plotted as functions of $ M/M_{\odot} $ using the best-fit values of the quantum correction parameter $ \xi $. The best-fit values of $\xi$ for these QPO sources are $\xi = 2.74 \pm 0.01$ for the upper frequency and $\xi = 2.68 \pm 0.03$ for the lower frequency. It is worth, however, noting that the errors in the best-fit values of $\xi$ result from the observational uncertainties incorporated into the fitting procedure (as seen in Table~\ref{Table 1}). It should be emphasized that these best-fit values do not imply a discovery of a non-zero $\xi$, since their determination is based on the simplifying assumption that QPOs are generated at the innermost stable circular orbit ($r_{\mathrm{ISCO}}$). In realistic scenarios, this assumption may not always hold. Nevertheless, the obtained best-fit values suggest the possible existence of a unique $\xi$ that remains consistent across those mentioned black hole candidates.

\renewcommand{\arraystretch}{1.3}
\begin{table}[]
\resizebox{1.0\textwidth}{!}{
\begin{tabular}{|c|c|c|c|c|}
\hline
        & GRS 1915 + 105 \cite{Remillard_2006} & H1743-322  \cite{Molla_2017,Ingram_2014}      & XTE J1550 - 564 \cite{Remillard_2002,Orosz_2011} & GRO J1655 - 40 \cite{mota2013} \\ \hline
$M [M_{\odot}]$    & $12.4^{+2.0}_{-1.8} $   & $11.21^{+1.65}_{-1.96} $  & $9.1^{+0.61}_{-0.61}$       & $5.4^{+0.3}_{-0.3}$        \\ \hline
$\nu_U [Hz]$  & $168^{+3.0}_{-3.0} $      & $240^{+3.0}_{-3.0} $         & $276^{+3.0}_{-3.0}$          & $441^{+2.0}_{-2.0}$         \\ \hline
$\nu_L [Hz]$ & $113^{+5.0}_{-5.0} $        & $165^{+9.0}_{-5.0}$         & $184^{+5.0}_{-5.0}$          & $298^{+4.0}_{-4.0}$         \\ \hline
\end{tabular}
}
 \caption{{The mass, upper, and lower frequencies of the QPOs from the X-ray binaries selected for analysis.}}
    \label{Table 1}
\end{table}
To estimate the limiting value of $\xi$ for each QPO source, we used the \textit{emcee} Python package \cite{Foreman_Mackey_2013}, which employs the Markov Chain Monte Carlo (MCMC) method to constrain the physical parameters. The posterior probability distribution, based on Bayes' theorem, is expressed as:
\begin{equation}
P(\Theta | D, \mathcal{M}) = \frac{L(D | \Theta, \mathcal{M}) \mathcal{P}(\Theta | \mathcal{M})}{P(D | \mathcal{M})} \, ,
\end{equation}
where $ P(\Theta | D, \mathcal{M}) $ represents the \textit{posterior probability}, 
$ L(D | \Theta, \mathcal{M}) $ is the \textit{likelihood function}, 
$ \mathcal{P}(\Theta | \mathcal{M}) $ is the \textit{prior probability}, 
and \( P(D | \mathcal{M}) \) serves as a \textit{normalizing constant}.
To specify the priors, we assume a Gaussian prior distribution for the model parameters within specified boundaries:

\begin{equation}
\mathcal{P}(\theta_i) \sim \exp \left[ \frac{1}{2} \left( \frac{\theta_i - \theta_{0,i}}{\sigma_i} \right)^2 \right], \quad \theta_{low,i} < \theta_i < \theta_{high,i} \, ,
\end{equation}
where \( \theta_i = [M, r, \xi] \) and \( \sigma_i \) are their corresponding standard deviations. The prior values for these parameters are adopted as presented in Table~\ref{table:priors}.

\renewcommand{\arraystretch}{1.2}
\begin{table}[]
\resizebox{1.0\textwidth}{!}{
\begin{tabular}{|c|cc|cc|cc|cc|}
\hline
\multirow{2}{*}{Parameters} & \multicolumn{2}{c|}{GRS 1915 + 105}   & \multicolumn{2}{c|}{H1743-322}        & \multicolumn{2}{c|}{XTE J1550 - 564}  & \multicolumn{2}{c|}{GRO J1655 - 40}   \\ \cline{2-9} 
                            & \multicolumn{1}{c|}{$\ \quad\mu\quad\ $} & $\sigma$ & \multicolumn{1}{c|}{$\ \quad\mu\quad\ $} & $\sigma$ & \multicolumn{1}{c|}{$\ \quad\mu\quad\ $} & $\sigma$ & \multicolumn{1}{c|}{$\ \quad\mu\quad\ $} & $\sigma$ \\ \hline
$M/M_{\odot}$               & \multicolumn{1}{c|}{12.4}  & 0.6      & \multicolumn{1}{c|}{11.21} & 0.6      & \multicolumn{1}{c|}{9.1}   & 0.6      & \multicolumn{1}{c|}{5.4}   & 0.3      \\ \hline
$r/M$                       & \multicolumn{1}{c|}{7.4}   & 0.3      & \multicolumn{1}{c|}{7.26}   & 0.3      & \multicolumn{1}{c|}{7.2}   & 0.3      & \multicolumn{1}{c|}{7.3}   & 0.3      \\ \hline
$\xi/M$                     & \multicolumn{2}{c|}{Uniform {[}0, 5)} & \multicolumn{2}{c|}{Uniform {[}0, 5)} & \multicolumn{2}{c|}{Uniform {[}0, 5)} & \multicolumn{2}{c|}{Uniform {[}0, 5)} \\ \hline
\end{tabular}
}
\caption{Gaussian priors for the selected X-ray binaries used in the MCMC analysis of QPOs occurring around QCBH~\cite{Liu:2023vfh}.}
\label{table:priors}
\end{table}

For the quantum correction parameter \( \xi \), we assume a uniform prior distribution over a specified range:

\begin{equation}
\mathcal{P}(\xi) = 1, \quad \text{for } \xi \in [0,5] \, ,
\end{equation}
otherwise, we set \( \mathcal{P}(\xi) = 0 \). This choice ensures that the parameter space remains constrained within a physically meaningful range.

Based on the upper and lower frequency expressions in Eqs.~\eqref{upper-lower-freq}, our MCMC analysis incorporates data from both the upper and lower frequencies (see Table~\ref{Table 1}). Consequently, the likelihood function $L$ is expressed as:
\begin{equation}
    \log L = \log L_{U} + \log L_{L} \, ,
\end{equation}
where $\log L_{U}$ denotes the likelihood of  the upper frequencies,
\begin{equation}
\log L_{U} = -\frac{1}{2} \sum_{i} \frac{(\nu^{i}_{U, obs} - \nu^{i}_{U, model})^2}{(\sigma^{i}_{U, obs})^2} \, ,
\end{equation}
and $\log L_{L}$ represents the likelihood of  the lower frequencies,
\begin{equation}
\log L_{L} = -\frac{1}{2} \sum_{i} \frac{(\nu^{i}_{L, obs} - \nu^{i}_{L, model})^2}{(\sigma^{i}_{L, obs})^2} \, .
\end{equation}
Here, $\nu^{i}_{U, obs}$ and $\nu^{i}_{L, obs}$ represent the measured values of the upper and lower frequencies, while $\nu^{i}_{U, model}$ and $\nu^{i}_{L, model}$ correspond to their predicted values from the theoretical model. The terms $\sigma^{i}_{U, obs}$ and $\sigma^{i}_{L, obs}$ represent the statistical uncertainties associated with the measured upper and lower frequencies, respectively.
Following the above setup, we perform an MCMC analysis to investigate the three-dimensional parameter space \{$r, M, \xi$\} associated with the quantum-corrected black hole (QCBH).  The best-fit values for these parameters are summarized in Table \ref{table:best-fit}, while Fig.~\ref{Fig.mcmc} presents the MCMC posterior distributions for four selected astronomical objects. The shaded regions in the Fig.~\ref{Fig.mcmc} correspond to the 68\%, 90\%, and 95\% confidence levels, offering insight into the statistical uncertainties of the inferred parameters.

The results indicate that the best-fit value of the quantum correction parameter $ \xi $ is obtained from GRO J1655-40, with an upper limit of $ \xi \leq 2.086 $ at the 95\% confidence level. GRO J1655-40 was selected due to its higher measurement accuracy compared to other QPO sources (see Table \ref{Table 1}). XTE J1550-564 and H1743-322 produced constraints close to that of GRO J1655-40, while GRS 1915+105 yielded a slightly higher upper limit. Therefore, we conclude that the quantum correction parameter is constrained to 
\begin{equation} \label{xi-qpo}
    \xi \leq 2.086 \ .
\end{equation}
In this study, we obtained an upper limit of $ \xi \leq 2.086 $, which provides a tighter constraint compared to the limit $ \xi \leq 2.304 $ obtained from the M87 black hole shadow \cite{shu2024qcbh} and the limit $ \xi \leq 2.866 $ derived from the Sgr A* shadow \cite{konoplya2024probingeffectivequantumgravity}. Unlike previous works that relied on black hole shadow observations to constrain $ \xi $, our approach utilizes QPO data from X-ray binaries, offering an alternative and complementary method for testing quantum corrected black holes in strong gravitational regimes.

\renewcommand{\arraystretch}{1.3}
\begin{table}[]
\resizebox{1.0\textwidth}{!}{
\begin{tabular}{|c|c|c|c|c|}
\hline
Parameters    & GRS 1915 + 105 & H1743-322       & XTE J1550 - 564 & GRO J1655 - 40 \\ \hline
$M/M_{\odot}$ & $12.45908_{-0.56419}^{+0.56642}$   & $10.80299_{-0.53946}^{+0.54902}$ & $8.81372_{-0.48624}^{+0.49328}$       & $5.26059_{-0.21233}^{+0.20339}$        \\ \hline
$r/M$         & $7.43238_{-0.28595}^{+0.28877}$        & $7.14127_{-0.27534}^{+0.27548}$         & $7.14602_{-0.27173}^{+0.27428}$          & $7.24254_{-0.24919}^{+0.25112}$         \\ \hline
$\xi/M$       & $<2.952$        & $<2.120$         & $<2.267$          & $<2.086$         \\ \hline
\end{tabular}
}
\caption{The best-fit values of QCBH parameters derived from QPOs in X-ray binaries.}
\label{table:best-fit}
\end{table}

\section{Conclusions}
\label{Sec:conclusion}

In this work, we aimed to constrain the quantum correction parameter $\xi$ for quantum corrected black holes (QCBHs) using QPO observations from X-ray binaries, along with the perihelion shifts of Mercury in the Solar System and the S2 star in $\text{SgrA}^\star$ SMBH environment.  
First, we analyzed the effect of the quantum correction parameter $\xi$ on the motion of a particle moving in the equatorial plane around the QCBH. We derived an analytical expression for the perihelion shift in the case of elliptical motion. Our results indicate that the perihelion shift decreases as the parameter $\xi$ increases (see Fig.~\ref{fig:geodesics}). By calculating the perihelion shifts of Mercury and the S2 star and comparing the results with observational data, we evaluated the upper bounds of $\xi$ to be $\xi \leq 0.01869$ for Mercury and $\xi \leq 0.73528$ for the S2 star.

Subsequently, we utilized QPOs to estimate the parameter $\xi$ in strong gravitational field regime conditions. We derived expressions for the radial ($\nu_r$) and latitudinal ($\nu_\theta$) frequencies of a test particle in the QCBH spacetime.
Our analysis showed that these frequencies decrease the parameter as $\xi$ increases (see Figs.~\ref{fig:frequency} and \ref{fig:freq-UP}). Using the Markov Chain Monte Carlo (MCMC) method, we constrained the limiting values of the parameter $\xi$ for four X-ray binaries. As a result, we obtained an upper limit of the parameter $\xi \leq 2.086$, providing tighter constraints compared to previous studies based on black hole shadows, such as those for M87 ($\xi \leq 2.304$) and Sgr A* ($\xi \leq 2.866$) \cite{shu2024qcbh,konoplya2024probingeffectivequantumgravity}.

These findings suggest that quantum gravitational effects are less significant in weak gravitational fields, where the parameter $\xi$ is smaller. In contrast, the influence of the quantum correction parameter becomes more pronounced in strong gravitational fields, as evidenced by the larger upper bound of the parameter $\xi$ obtained from QPO observations.

Our results provide new insights into the role of the quantum correction parameter in quantum corrected black holes. Future high-precision observations, particularly with next-generation telescopes, are expected to refine these constraints further, advancing our understanding of quantum gravitational phenomena.

\acknowledgments

We are thankful to the anonymous referee for constructive suggestions and comments that helped us improve the clarity and quality of the manuscript. S.S. is supported by the National Natural Science Foundation of China under Grant No. W2433018. T.Z. is also supported by the National Natural Science Foundation of China under Grants No. 12275238, the National Key Research and Development Program of China under Grant No. 2020YFC2201503, and the Zhejiang Provincial Natural Science Foundation of China under Grants No. LR21A050001 and No. LY20A050002.




%
\bibliographystyle{apsrev4-1}  
\bibliography{Ref1,Ref2}

\end{document}